\documentclass[aip,preprint]{revtex4-1}
\usepackage{graphicx}
\usepackage{amsmath}
\usepackage{siunitx}
\usepackage{xcolor}

\draft 

\begin{document}

\title{100-kT Magnetic field generation using paisley targets by femtosecond laser-plasma interactions}
\author{M-A.H. Zosa}
\affiliation{Division of Electrical, Electronic and Infocommunications Engineering, Graduate School of Engineering, Osaka University, 2-1 Yamadaoka, Suita, Osaka 565-0871, Japan}
\affiliation{%
Institute of Laser Engineering, Osaka University, 2-6 Yamadaoka, Suita, Osaka 565-0871, Japan}
\author{Y.J. Gu}%

\author{M. Murakami}
\email{murakami-m@ile.osaka-u.ac.jp}
\affiliation{%
Institute of Laser Engineering, Osaka University, 2-6 Yamadaoka, Suita, Osaka 565-0871, Japan}

\date{\today}

\begin{abstract}
A target using a paisley pattern generates 100-\si{\kilo\tesla}-level magnetic fields. Laser irradiation induces local charge separation on the target, which creates surface currents along the concave surface, generating a magnetic field. For a laser intensity of $10^{21}$\si{\watt\per\square\centi\meter}, the target generates a 150-\si{\kilo\tesla} magnetic field. We developed a simple model to describe the magnetic field as a function of laser intensity and target radius. A double paisley configuration extends the lifetime of the magnetic field to the picosecond scale. The paisley design generates comparable results even if it is simplified. Thus, it is a robust and modular target suitable for magnetic field applications such as 100-kT magnetic field generation and magnetic reconnection.
\end{abstract}

\maketitle

Modern developments in relativistic femtosecond lasers\cite{Yoon2019,Yoon2021} and microstructure fabrication\cite{Fritzler2019,Fu2016} have expanded the scope of high energy density physics\cite{Snyder2019,Zhou2021,Sugimoto2020}. Recently, studies have utilized these developments to investigate ion acceleration\cite{Adli2019}, magnetic field generation\cite{Jiang2021}, ultra-high density compression\cite{Murakami2018,Murakami2019}, and pair-creation\cite{Koga2020}. Generating magnetic fields on the 100-kT scale is exciting because it enables the study of fundamental phenomena such as magnetic reconnection\cite{Fujioka2013,Gu2021,Pei2016}. Magnetic fields on this scale are observed in the accretion disks of black holes, which makes them valuable for laboratory astrophysics experiments\cite{Law2020}. 

Irradiating an ``escargot" target with a laser is a well-known scheme to produce a strong magnetic field\cite{Korneev2015,Korneev2017}. It has been used in a laboratory experiment as a magnetic field source\cite{Law2020}. Microtube implosion (MTI) is another method for magnetic field generation\cite{Murakami2020,Weichman2020}. In MTI, the implosion of the inner layer of a microtube amplifies a seed magnetic field to the megatesla scale, enhancing its strength by a few orders of magnitude\cite{Shokov2021}.

\begin{figure}[htbp]
    \centering
    \includegraphics{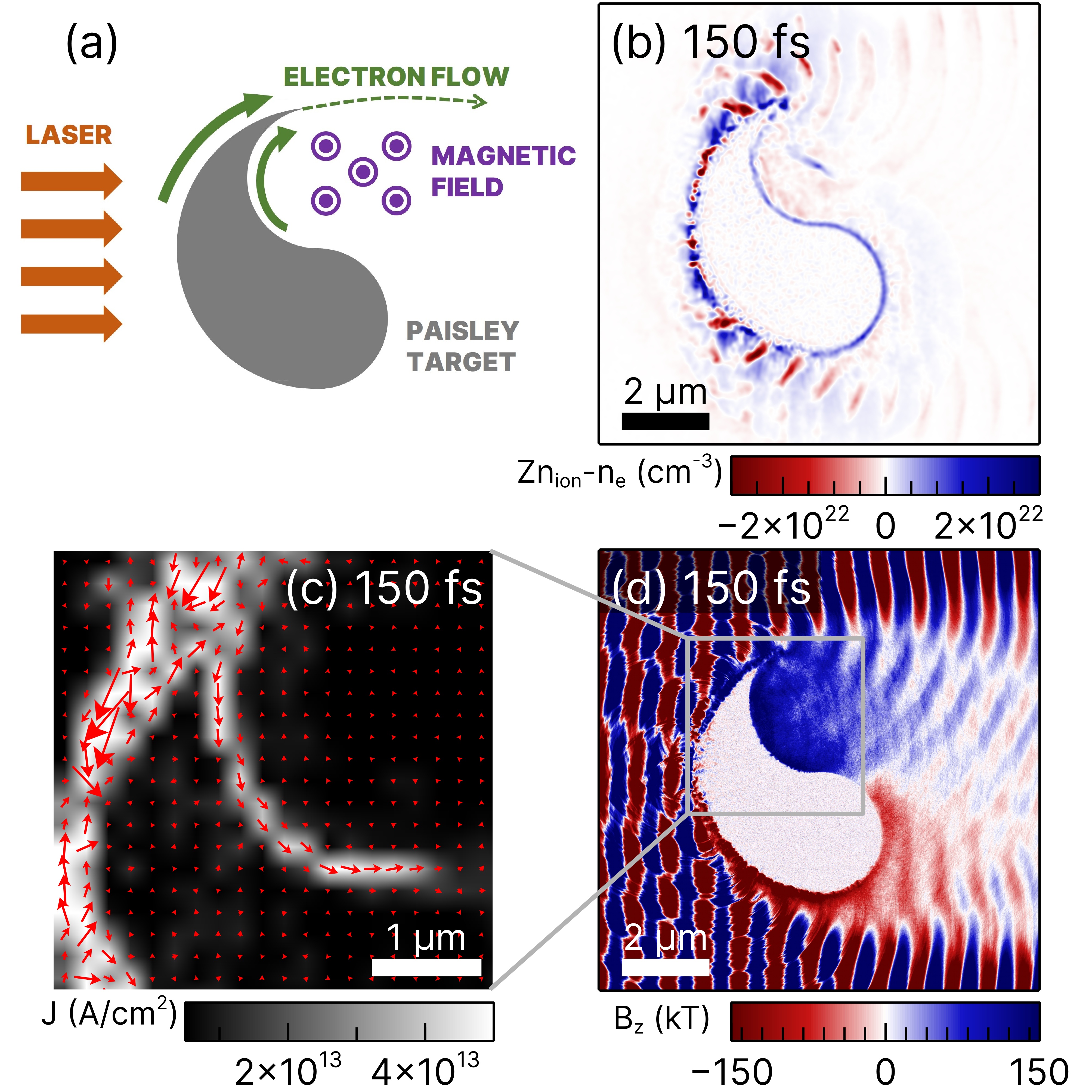}
    \caption{(a) Illustration of the magnetic field generation mechanism. At t = \SI{150}{\femto\second}, (b) charge separation profile, (c) vector diagram for the net current, and (d) magnetic field profile along the z-axis. }
    \label{fig:single}
\end{figure}

Although both setups utilize different approaches to generate a magnetic field, an essential factor is the formation of a surface current\cite{Weichman2020_2,Korneev2015,Weichman2020}. In this work, we propose a paisley design to generate a magnetic field without a seed magnetic field. Our paisley design is described mathematically by the following function
\begin{equation}
f(k) = \begin{cases} 
      -\frac{R_0}{2}\left(\exp{\left[4\pi ik - \pi/2 \right]} + 1\right) & k \in \left(0, \frac{1}{4} \right) \\
      -\frac{R_0}{2}\left(\exp{\left[4\pi ik - \pi/2 \right]} - 1\right)  & k \in \left(\frac{1}{4}, \frac{1}{2} \right) \\
      R_0\exp{\left(2\pi ik - \pi/2 \right)}  & k \in \left(\frac{1}{2},1 \right) \label{eq:paisley}
   \end{cases},
\end{equation}
where $R_0$ is the radius, and $k\in \left(0,1 \right)$ is a parametric variable. The x- and y-coordinates are the real and imaginary parts of eq.~\eqref{eq:paisley}, respectively. Figure~\ref{fig:single}(a) graphically depicts Eq.~\eqref{eq:paisley}. 

In this design, surface currents produce a magnetic field on the concave side of the target, which makes the magnetic field easily accessible. The open area makes it easier for incoming particles to interact with the magnetic field. Additionally, the accessible location permits two or more targets to be connected in a modular fashion allowing the generated magnetic fields to interact with each other. Thus, it is suitable in experiments requiring the interaction of two or more magnetic field sources. Various arrangements can be used to study magnetic field phenomena such as magnetic reconnection, magnetic mirrors, and other laboratory astrophysics experiments. 

To study the magnetic field generation of the paisley design, we used the 2.5D particle-in-cell (PIC) program, EPOCH\cite{Arber:2015hc}. The laser parameters were $\lambda_L =\SI{800}{\nano\meter},\ I_L = \SI{1e21}{\watt\per\square\centi\meter}$, and $\tau_L = \SI{100}{\femto\second}$ for the wavelength, peak intensity, and full-width at half-maximum (FWHM), respectively. The simulations used a $\SI{30}{\micro\meter}\times \SI{30}{\micro\meter}$ box with a cell size of $\frac{\lambda_L}{100}$, $100$ pseudo-ions and $200$ pseudo-electron per cell, and a laser propagating in the +x direction. The target consisted of fully ionized carbon with a density of $\SI{1e23}{\per\cubic\centi\meter}$.

The paisley target generates a surface current via local charge separation. The thickness gradient along the tip creates a larger charge separation around the apex upon laser irradiation [Fig.~\ref{fig:single}(b)]. The laser strips most of the electrons from the thin sections of the paisley target, but it cannot penetrate the thicker areas. Hence, more electrons are ejected close to the apex. This causes the surface electrons to flow towards the apex [Fig.~\ref{fig:single}(c)]. The curvature of the surface causes a positive (negative) magnetic field to form on its concave (convex) side [Fig.~\ref{fig:single}(d)]. Although the magnetic field covers only a few square microns, using a larger target will increase its coverage. When using larger targets, materials with low electron densities, such as foam, are preferred because low-density materials have a larger skin depth. The larger skin depth enables the larger target to maintain the charge separation gradient across the tip. Additionally, if the target is rotated $180^{\circ}$ about $y=0$, the polarity of the magnetic fields flips.

\begin{figure}[htbp]
    \centering
    \includegraphics{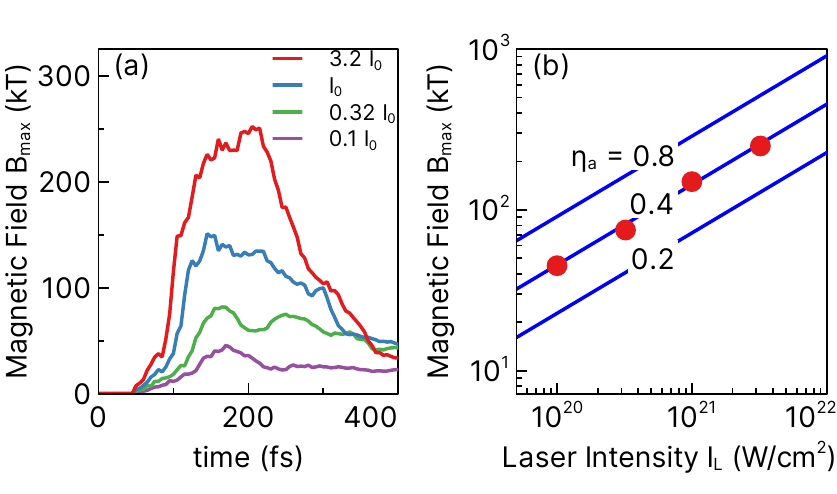}
    \caption{(a) Maximum magnetic field generated by the paisley target vs time for various laser intensities $(I_0 = 10^{21}\si{\watt\per\square\centi\meter})$. (b) Simple analytic model (blue lines) for varying $\eta_a$ and the maximum magnetic field from the simulations results (red dots).}
    \label{fig:model}
\end{figure}

To predict magnitude of the magnetic field in \si{\kilo\tesla}, we developed a simple analytic model. The magnetic field strength is $B_z\sim j_e R_0$, and the estimated current density, $j_e$, is $j_e\sim n_{he} c$. The hot-electron density, $n_{he}$, is related to $I_L$ by $n_{he} = {\eta_a I_L}/{\mathcal{E}c}$\cite{Forslund1977}. For relativistic electrons, the average kinetic energy, $\mathcal{E}$, is approximately $3T_e$, where $T_e$ is the electron temperature. If the electron temperature is estimated using the ponderomotive scaling\cite{Wilks1992}, the model is reduced to
\begin{equation}
    B_z = 30.3 \frac{\eta_a\sqrt{I_{L20}}R_{0\mu m}}{\lambda_{\mu m}},\label{eq:model}
\end{equation}
where $I_{L20}$ is the laser intensity normalized to $10^{20}$~\si{\watt\per\square\centi\meter}, $\eta_a$ is the absorption efficiency, $R_{0 \mu m}$ is characteristic radius given by Eq.~\eqref{eq:paisley} in \si{\mu m}, and $\lambda_{\mu m}$ is the laser wavelength in \si{\micro\meter}. The magnetic field strength scales as $B_z \sim \sqrt{I_L}$, and linearly with $R_0$. Figure~\ref{fig:model}(a) shows that the peak magnetic field increases with the laser intensity. The FWHM for the magnetic field is $\sim 2\tau_L$. Figure~\ref{fig:model}(b) shows that the PIC simulation results agree well with Eq.~\eqref{eq:model} for $\eta_a = 0.4$. According to the model, an absorption efficiency of $0.8$ or higher is necessary to reach the megatesla scale for $I_L = 10^{22}~\si{\watt\per\square\centi\meter}$. However, at this intensity, the model may inaccurate because non-linear effects are no longer be negligible.

\begin{figure*}[htbp]
    \centering
    \includegraphics{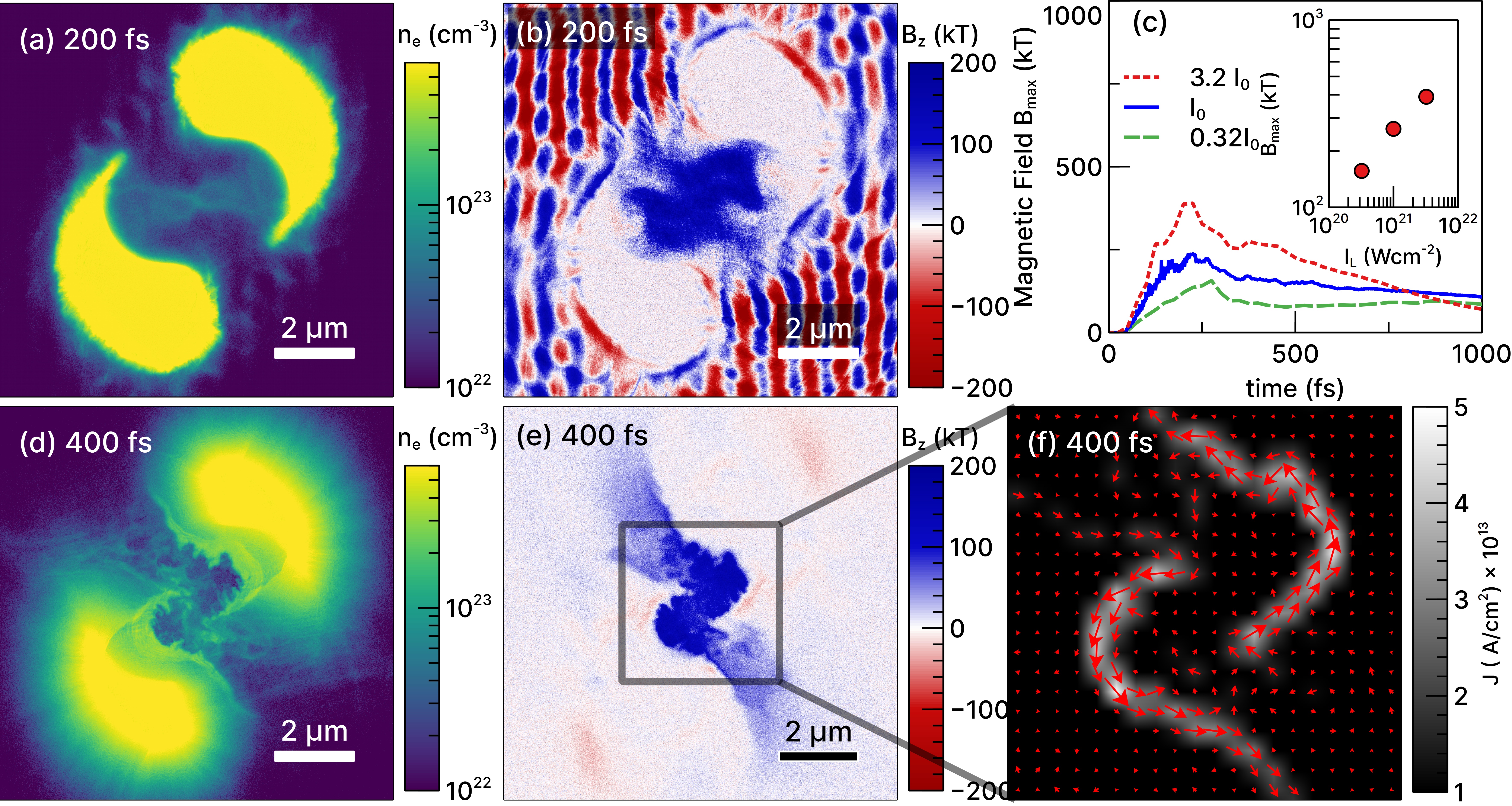}
    \caption{At t = \SI{200}{\femto\second}, (a) the electron density profile and (b) the magnetic field profile, $B_z$. (c) Maximum magnetic field as a function of time at different intensities. The inset shows the magnetic field strength as a function of laser intensity. At t = \SI{400}{\femto\second} (d) the electron density profile, (e) the magnetic field profile, and (f) the current vector diagram.}
    \label{fig:double}
\end{figure*}

Comparing Fig.~\ref{fig:model}(a) with the simulation parameters, the peak magnetic field coincides with the laser maximum. Additionally, the magnetic field sharply drops once the laser stops interacting with the target. This results in a relatively short magnetic field lifetime. However, using two paisley targets prolongs the magnetic field lifetime [Fig.~\ref{fig:double}(c)]. In this case, the two lasers hit a pair of paisley targets from the +x and --x-direction. The two targets are separated by a gap to minimize the possibility of electrons flowing directly from the body of one target to the tip of the other. The magnetic field generated by the double paisley setup almost completely covers the void [Figs.~\ref{fig:double}(a) \& (b)]. Additionally, the magnetic field is sustained for much longer than the laser pulse duration. As the system evolves, electrons flow towards the center [Fig.~\ref{fig:double}(d)] and form a partial current loop [Fig.~\ref{fig:double}(e)]. This loop is stable and extends the magnetic field's lifetime to the picosecond scale [Fig.~\ref{fig:double}(f)]. Although the maximum magnetic field strength has a long lifetime, Fig.~\ref{fig:double}(e) shows that the magnetic field leaks from the confined space. This results in a gradual reduction of the total magnetized area. By \SI{1}{\pico\second}, the 100-kT region is estimated to be 20\% of the area at \SI{400}{\femto\second}, and the magnetic field is reduced by one order of magnitude by \SI{2}{\pico\second}.  Figure~\ref{fig:single}(b) shows that a positive patch forms on the concave-side. It is attributed to the imploding ions. The imploded ions attract electrons whose trajectories are bent by the magnetic field generated by the surface current\cite{Gu2021_2}. The electron gyro motion around the imploded ions works to sustain the magnetic field for a brief period after the laser has disappeared in the single paisley case. For the double paisley targets, the imploded ion region is more pronounced, which helps sustain a partial current loop [Fig.~\ref{fig:double}(f)]. In addition, two paisley targets form a more confined region, which delays the expansion of the current loop. The electron collision frequency is calculated using a simple formula\cite{Weichman2020_2}. This formula gives the characteristic time scale of the electron collision , $\nu_e^{-1}$, which is several picoseconds long. Thus, dissipation due to Coulomb collisions is negligible for the duration of the magnetic field. For comparison, we also conducted simulations of a \SI{5}{\micro\meter} ``escargot" target. It generates a magnetic field of \SI{150}{\kilo\tesla} with a picosecond lifetime using the same laser parameters.

\begin{figure*}[htbp]
    \centering
    \includegraphics{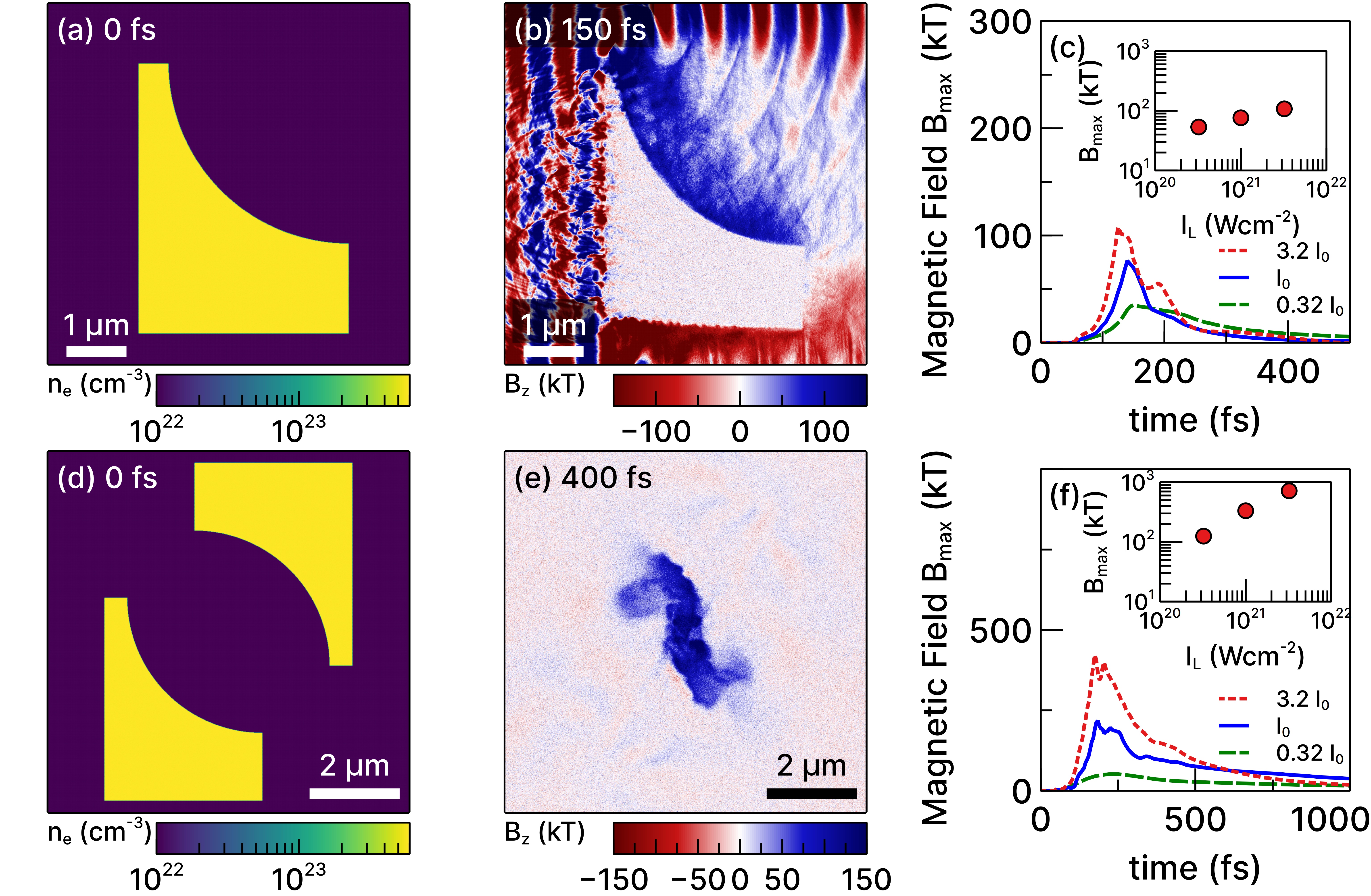}
    \caption{Simplified single paisley design: (a) Initial electron density profile, (b) magnetic field profile at $t=\SI{150}{\femto\second}$, and (c) time evolution of the magnetic field strength. Inset shows the magnetic field strength as a function of laser intensity. Simplified double paisley design: (d) Initial electron density profile and (e) magnetic field profile at $t=\SI{400}{\femto\second}$, and (f) time evolution of the magnetic field strength. Inset shows the magnetic field strength as a function of laser intensity.}
    \label{fig:simp}
\end{figure*}

A drawback of the paisley design is its intricate shape, which is challenging to fabricate. However, a simplified design will yield comparable results. Figure~\ref{fig:simp}(a), shows that a quarter of a rectangular microtube can be used as a simplified paisley target. Although the design differs from the original one, the core concept of utilizing the thickness gradient to guide the surface current remains. Despite a major change in its appearance, the magnetic field strength produced by the simplified target is comparable to the paisley design [Fig.~\ref{fig:simp}]. However, it has a slightly smaller cross-section, and a shorter lifetime. Although the simplified design might be easier to fabricate, the original paisley structure is still interesting from a theoretical viewpoint.

\begin{figure}[htbp]
    \centering
    \includegraphics{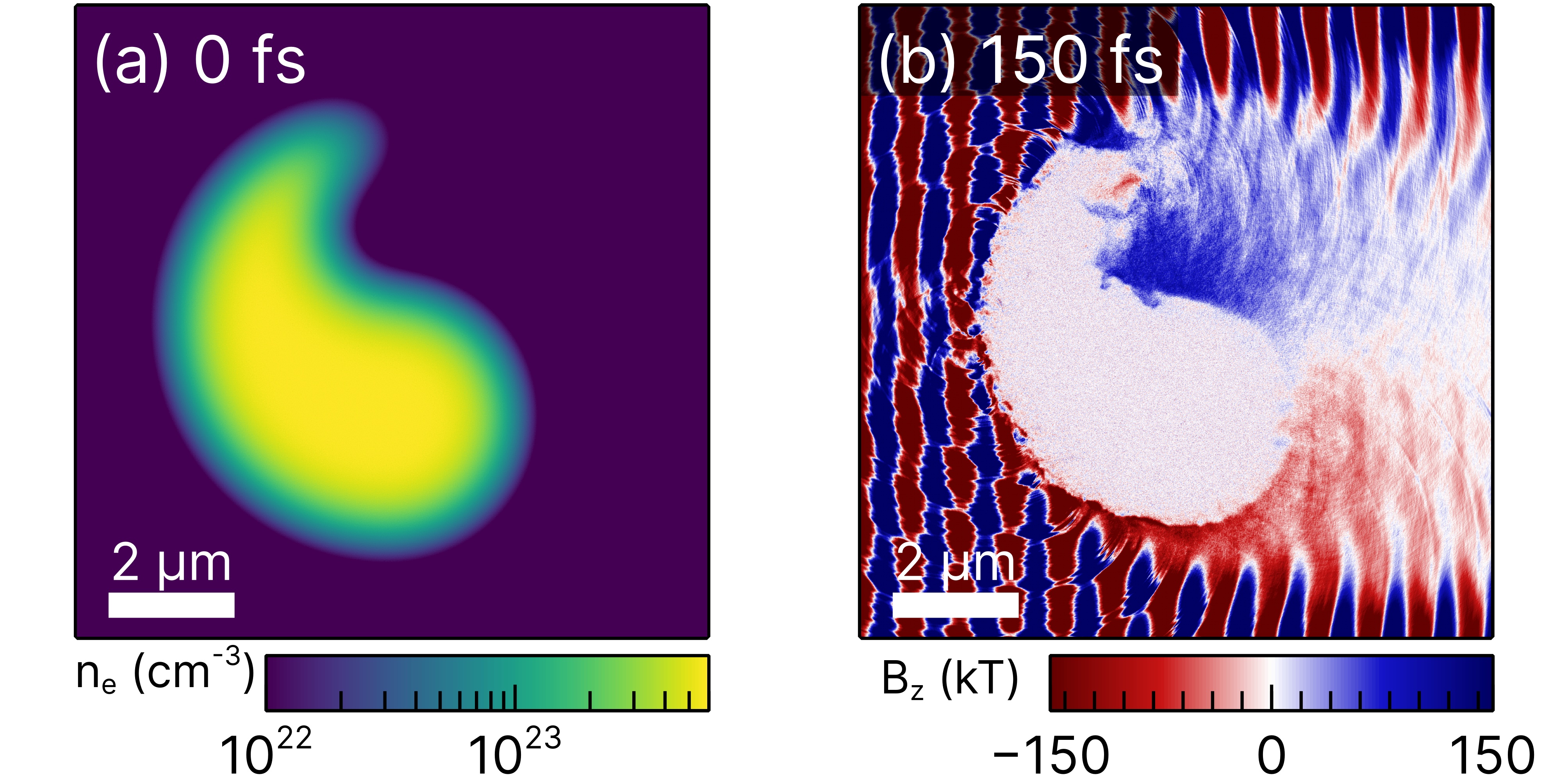}
    \caption{(a) Initial electron density profile and (b) magnetic field profile at $t=\SI{150}{\femto\second}$ for the pre-expanded case.}
    \label{fig:pre}
\end{figure}

Due to its small size, the paisley target is prone to pre-expansion when interacting with the laser's pre-pulse. To approximate this effect, we modified the initial plasma distribution profile of the paisley target. Figure~\ref{fig:pre} shows the simulation results of a paisley target with a modified initial density profile [Fig.~\ref{fig:pre}(a)]. Although the area of the magnetic field in Fig.~\ref{fig:pre}(b) is smaller than that in Fig.~\ref{fig:single}(d), the results show that even when the initial distribution is not ideal, it can still produce 100-kT magnetic fields. The potential 3D-effects is another factor to consider for these targets. The influence of 3D effects should be more prominent on the top and bottom (z-axis) ends of the target because the z-axis expansion on the ends alters the electron dynamics close to the ends. This effect can be mitigated by choosing a relatively long or high aspect-ratio target. For experimental verification, the magnetic field strength can be evaluated by measuring the deflection of a passing ion beam\cite{Law2020}.

The paisley target is a robust design to generate a magnetic field without a seed. However, further optimization can still be performed to maximize the generated magnetic field. It has potential for multiple applications due to its flexible and modular design. For the double paisley target, its performance is similar to the ``escargot" target for both the magnetic field intensity and lifetime. However, the double paisley target requires two lasers, which is less efficient than the ``escargot" target. Nevertheless, the paisley target's advantage lies in its flexibility. Although the current double paisley setup is used to prolong the magnetic field lifetime, flipping one of the paisley targets will result in two magnetic fields with opposing polarities. This configuration is suitable for studying magnetic reconnection. Additionally, different configurations of the paisley targets may be realized to study other magnetic field interactions.

\begin{acknowledgments}
The authors acknowledge Didar Shokov for the fruitful discussions. Computational resources were provided by the Cybermedia Center, Osaka University. This work was supported by the 
Japan Society for the Promotion of Science (JSPS). PIC simulations were performed using EPOCH, developed under UK EPSRC (Grant Nos. EP/G054940, EP/G055165, and EP/G056803).
\end{acknowledgments}

\section*{Data Availability}
The data that support the findings of this study are available from the corresponding author upon reasonable request.

\section*{Author Declarations}
\subsection*{Conflict of Interest}
The authors have no conflicts to disclose.
\bibliography{Ref.bib}
\newpage

\end{document}